\documentclass[twocolumn,secnumarabic,amssymb, nofootinbib, aps, prd, superscriptaddress]{revtex4-1}
\usepackage{amsmath}
\usepackage{graphicx}
\usepackage{float}
\usepackage{tabularx}
\usepackage{color}
\usepackage[dvipsnames]{xcolor}
\usepackage{parskip}
\usepackage{xfrac}
\usepackage{lineno}
\usepackage[normalem]{ulem}
\usepackage[colorlinks,urlcolor=NavyBlue,citecolor=NavyBlue,linkcolor=NavyBlue,pdfusetitle]{hyperref}

\begin{document}

\title{Machine-learning non-stationary noise out of gravitational-wave detectors}

\date{\today}

\author{G. Vajente}\email{vajente@caltech.edu}\affiliation{California Institute of Technology, \\ LIGO Laboratory MC 100-36, \\ 1200 E. California Blvd. Pasadena (CA) USA}
\author{Y. Huang}\affiliation{Massachusetts Institute of Technology, \\ Kavli Institute for Astrophysics and Space Research, \\ 77 Massachusetts Avenue, Cambridge (MA) }
\author{M. Isi}\affiliation{Massachusetts Institute of Technology, \\ Kavli Institute for Astrophysics and Space Research, \\ 77 Massachusetts Avenue, Cambridge (MA) }
\author{J. C. Driggers}\affiliation{LIGO Hanford Observatory, Richland, WA 99352, USA }
\author{J. S. Kissel}\affiliation{LIGO Hanford Observatory, Richland, WA 99352, USA }
\author{M.~J.~Szczepa\'nczyk}\affiliation{University of Florida, Gainesville, FL 32611, USA}
\author{S. Vitale}\affiliation{Massachusetts Institute of Technology, \\ Kavli Institute for Astrophysics and Space Research, \\ 77 Massachusetts Avenue, Cambridge (MA) }

\begin{abstract}
Signal extraction out of background noise is a common challenge in high-precision physics experiments, where the measurement output is often a continuous data stream.  To improve the signal-to-noise ratio of the detection, witness sensors are often used to independently measure background noises and subtract them from the main signal. If the noise coupling is linear and stationary, optimal techniques already exist and are routinely implemented in many experiments. However, when the noise coupling is non-stationary, linear techniques often fail or are sub-optimal. Inspired by the properties of the background noise in gravitational wave detectors, this work develops a novel algorithm to efficiently characterize and remove non-stationary noise couplings, provided there exist witnesses of the noise source and of the modulation. In this work, the algorithm is described in its most general formulation, and its efficiency is demonstrated with examples from the data of the Advanced LIGO gravitational-wave observatory, where we could obtain an improvement of the detector gravitational-wave reach without introducing any bias on the source parameter estimation. \end{abstract}

\maketitle

\section{Introduction}\label{sec:introduction}


High-precision measurements in physics rely on the ability to separate interesting signals from background noise. In many modern experiments, the instrument output is a continuous stream of data, and signal processing techniques have been developed to characterize and remove noise from data streams. In the simplest possible case, the disturbance can be modeled as an additive noise having constant statistical properties (for example, power spectral density) over time. This is the case of \emph{stationary noise}: most signal detection techniques have been developed under this assumption, and are optimal when the noise is stationary and gaussian. Additionally, if the noise can be probed by additional witness sensors, which are known to be insensitive to the targeted signal, there exist many techniques to efficiently subtract the noise from the main signal, thus improving the detection chances. In the linear and stationary noise coupling case, the optimal strategy is the Wiener filter \cite{wiener}.

In real world physical systems however, the noise is rarely stationary: the statistical properties can vary over time during the measurement. When this is the case, the signal detection algorithms that were optimal for stationary noise become sub-optimal, and might even be fooled by noise transients. The noise can still be sampled by auxiliary witness sensors, but the coupling from those witnesses to the main signal might be non-linear or non-stationary. In this case, noise cancellation techniques like the Wiener filter are not optimal or might fail altogether. 

The distinction between a non-linear and a non-stationary noise coupling is simply a matter of time scales or frequencies. Consider, for example, two auxiliary signals $x(t)$ and $y(t)$ that couple into the main detector output $d(t)$ as the product $d(t) = x(t) \cdot y(t)$. If both signals contain significant power in the frequency range of interest for the measurement being performed, then the noise coupling manifests itself as non-linear, since there is never any linear relationship between the individual noise witnesses and the detector output. However, if one of the two signals $x$ has power only at very low frequencies, then for periods of time shorter than the typical time scales that characterize the variation of $x$, the coupling of $y$ to $d$ is linear and approximately constant. In this case, we would consider the noise coupling to be linear, but modulated in time. A possible approach to the subtraction of this \emph{non-stationary noise coupling} is to use adaptive filtering techniques \cite{adaptive}. Instead, this work develops a more efficient solution, which is applicable when the noise coupling modulation is sensed by any number of witness channels, i.e. when the source of the modulation is measurable.

The work presented here is of general applicability to signal processing, although  inspired by work on gravitational wave interferometric detectors \cite{aligo, avirgo, kagra, geo600}. The now numerous detections of gravitational wave (GW) signals from the coalescence of binary systems \cite{gw-catalog} have opened the era of GW astronomy. The detection rate and the accuracy of the astrophysical inference about the source parameters and populations are strongly dependent on the detector sensitivity. Ideally, the sensitivity of a GW detector is limited by fundamental noise sources, such as quantum noise \cite{quantum-noise}, thermal noise \cite{thermal-noise} or gravity gradient noises \cite{newtonian-noise}. Real world instruments \cite{aligo, avirgo, kagra, geo600} are rarely limited only by fundamental noises, but rather by other, technical noises \cite{aligo-sensitivity} that are a consequence, for example, of the feedback control systems needed to maintain the correct operating point, or of unmodeled dynamical behavior of the apparatus. While fundamental noises are expected to be stationary, i.e. to have constant statistical properties over time, there is no reason to assume the same to be true for technical noises. Similarly, the coupling of noise sources from auxiliary degrees of freedom can contain non-linear terms beside the usually dominant linear contributions. In this case, the detector noise might look stationary on timescales longer than the non-linear dynamics timescale, but its statistics can be highly non-Gaussian. 

The presence of non-stationary noise can be problematic in different ways. First of all, fluctuations of the detector noise over short time scales (of the order of a second) can mimic transient GW signals and contaminate the data \cite{detchar-glitches}. Furthermore, many detection pipelines \cite{pycbc, gstlal} use matched filtering \cite{match-filtering}, which is optimal only when the background noise is Gaussian and stationary. The estimation of the significance of GW candidates can therefore suffer from the presence of noise that deviates from this assumption.

The main result of this work is an algorithm that can be used to characterize non-stationary noise couplings from multiple witness signals, and to subtract in the time domain the noise from a target signal, extending well-known techniques already used in the linear and stationary case \cite{wiener, driggers2018, davis2018}. This algorithm is able to model noise coupling modulations that are sensed by slowly-varying witness sensors, using an efficient parametrization that allows a time domain subtraction, free of unstable filters and over-fitting problems. This algorithm can also be applied to linear and stationary couplings, providing means to implement parametric and stable noise subtraction: this is therefore a viable approach to solve the problem of fitting and implementing time-domain Infinite Impulse Response (IIR) Wiener filters  \cite{wiener-iir}.

The rest of this article is organized as follows. Section \ref{sec:noise} describes non-linear and non-stationary noise couplings, and lays the basis for the mathematical description of the algorithm, which is then described in section \ref{sec:parametrized}. In section \ref{sec:results}, as an example application, the algorithm is applied to the Advanced LIGO GW detectors. It is worth noting that the non-stationary noise subtraction of the 60 Hz power line (described in section \ref{sec:60hz}) has already been implemented successfully in the Advanced LIGO detectors during the third observation run O3. Finally, section \ref{sec:extra} describes extensions and additional applications of this algorithm, and section \ref{sec:conclusions} concludes with final remarks and discussion.

\section{Non-linear and non-stationary noise couplings}\label{sec:noise}

From this point on we will discuss non-stationary noise couplings by considering the example of a gravitational wave detector output $h(t)$, but the discussion presented here is valid in general for any physical measurement system that provides a continuous data stream as an output. The detector output is the sum of real GW signals $h_{\mathrm{GW}}(t)$ and background noise, the latter can be subdivided into diverse contributions: fundamental noises $\varepsilon_\mathrm{F}(t)$ that cannot be measured or subtracted (like quantum noise or thermal noise); noises $\varepsilon_\mathrm{L}(t)$ that couple with a linear and time-stationary transfer function from auxiliary degrees of freedom and that can therefore be measured and subtracted; noises $\varepsilon_{\mathrm{NL}}(t)$ that couple from auxiliary degrees of freedom or channels in a non-linear or non-time-stationary way; finally there can be unknown noise sources $\varepsilon_\mathrm{U}(t)$ whose origin is not yet understood and that can not be measured in any other available channel.

In this section we focus on the case of non-linear or non-stationary noise couplings. Linear and stationary noise couplings will emerge as a special case of this treatment. We assume that the noise source can be monitored by a set of witness signals $w_i(t)$ with $i=1 \dots N$. We then model the detector output $h(t)$ as the sum of an uncorrelated and untrackable noise background $\varepsilon_\mathrm{B}(t)$ and the non-linear contribution related to the witness signals:
\begin{equation}\label{eq:full-non-linear}
h(t) = \varepsilon_\mathrm{B}(t) + \mathcal{F}\left[ w_1(\tau < t), \dots, w_N(\tau < t)\right]
\end{equation}
In this expression we already included two assumptions: causality, meaning that the contribution at the time $t$ can depend only on the witness values in the past; time invariance, expressed by the requirement that the functional form $\mathcal{F}\left[\cdot\right]$ does not contain any explicit dependency on $t$, meaning that all the time variation of the noise is encoded in the witness signals (we shall see in section \ref{sec:extra} how this requirement can be relaxed). We are given the detector output $h(t)$ and the witness sensors $w_i(t)$, and the task to find a suitable representation of the functional $\mathcal{F}$ so that we can optimally subtract the excess noise from $h(t)$. While in the case of linear coupling there are simple and efficient ways to parametrize the functional $\mathcal{F}$, such as a frequency- or time-domain Wiener filter \cite{wiener}, such general parametrization does not exist in the non-linear case. 

One possible solution to the parametrization problem is to use deep neural networks (DNN) \cite{dnn}, which have been proven to perform as universal function approximators, provided they are composed of layers with a large number of neurons \cite{dnn-approximators}. This approach was initially applied to the Advanced LIGO data, without satisfactory results, and is described in section \ref{sec:deep-learning}. The main drawback of using a DNN is its high complexity, which in turn causes a long training time, sub-optimal performances and  difficulty in interpreting the results \cite{black-box}.

The approach used in this work is inspired by common Machine Learning algorithms, but one of its key features is a large reduction of the model complexity (fewer parameters), achieved by adopting a model of the non-linear or non-stationary noise coupling. The model is potentially not as general as a DNN, but in all the cases we considered in the context of GW detection, it outperformed the DNN approach, due to the ease of training and interpretability of the results.

\section{Non-stationary parametrized subtraction}\label{sec:parametrized}

The most common form of non-linear coupling found in GW detectors consist of one "fast" noise source $n(t)$ that couples to the detector output through a linear transfer function, which is however ``slowly" changing over time, and this change can be tracked by additional ``slow" witness signals $w_i(t)$. The distinction between fast vs. slow will be explained below precisely. In brief it refers to the frequency content of the signals: the noise is relevant for the detector sensitivity at high frequencies (above 10 Hz), while the typical coupling modulation happens at lower frequencies (below 1 Hz). In this case, it is possible to describe the non-linear coupling with a truncated series expansion, where the different time scales can be separated. Each term in the series can then be parametrized in an efficient way and a numerical optimization algorithm used to minimize the impact of the noise in the target signal. This section explains this algorithm in detail.

The most general non-linear coupling, described in equation \ref{eq:full-non-linear}, can be expanded in a Volterra series \cite{volterra}, subdividing the non-linear terms in increasing polynomial orders. Restricting to the second order we can write:
\begin{eqnarray}\label{eq:volterra}
\varepsilon_\mathrm{NL}(t) &=& \mathcal{F}\left[ w_1(\tau < t), \dots, w_N(\tau < t)\right] \nonumber \\
&=& \sum_{i,j=1}^{N} \iint\limits_{0}^{\quad+\infty} \alpha_{i,j}(\tau_1, \tau_2) w_i(t - \tau_1) w_j(t -  \tau_2) \, d\tau_1 d\tau_2  \nonumber \\
 && + \dots 
\end{eqnarray}
where $\alpha_{i,j}$ are the second order Volterra kernels. It is useful to write the frequency domain equivalent of the expression above, by defining the Fourier transform of the kernels as:
\begin{equation}
\tilde \alpha_{ij}(\omega_1, \omega_2) = \iint\limits_{-\infty}^{\quad+\infty} \alpha_{ij}(\tau_1, \tau_2) e^{i\omega_1 \tau_1} e^{i\omega_2 \tau_2} \, d\tau_1 d\tau_2
\end{equation}
If we now substitute the inverse of this expression into the Volterra series, we find:
\begin{eqnarray}\label{eq:volterra-fdomain}
\tilde \varepsilon_\mathrm{NL}(\omega_3) &=&  \sum_{i,j=1}^{N} \iint\limits_{0}^{\quad+\infty} \delta(\omega_3 - \omega_1 - \omega_2) \cdot \nonumber \\
&& \tilde \alpha_{i,j}(\omega_1, \omega_2) \tilde w_i(\omega_1) \tilde w_j(\omega_2) d\omega_1 d\omega_2
\end{eqnarray}
where the tilde denotes the Fourier transform of a signal. This frequency-domain expression makes it clear that the quadratic term mixes the two input signal frequencies into the sum frequency in the target signal $\omega_3 = \omega_1 + \omega_2$. To simplify this expression we make a few important assumptions, splitting the set of all noise witnesses $\left\{w_i\right\}$ into two classes: one fast noise witness $n(t)$ and a set of slow modulation witnesses $x_i$. The first assumption is that the frequencies at which the noise source $\tilde n(\omega_1)$ is relevant for the detector output is much higher than the typical frequencies where the modulation witness signals $\tilde x_i(\omega_2)$ are concentrated. Typically, for a GW detector, the noise frequency of interest $\omega_1$ is in the 10 to 1000 Hz range, while the modulation signals are concentrated at frequencies $\omega_2$ below 1 Hz, so the assumption $\omega_1 \gg \omega_2$ is reasonable in the cases under consideration. This allows us to ignore the dependency of the Volterra kernels on the lower frequency $\omega_2$ and write $\tilde \alpha_{ij}(\omega_1, \omega_2) \simeq \tilde \alpha_{ij}(\omega_1)$. By transforming back to the time domain we find the expression below for the non-stationary noise coupling
\begin{equation}\label{eq:modulation-model}
\varepsilon_\mathrm{NL}(t) = \sum_{i=1}^N \int\limits_{0}^{+\infty} \alpha_i(\tau) n_i(t-\tau) \, d\tau
\end{equation} 
where each $n_i(t)$ is the time-domain product of the noise source with one of the modulation witness signals $n_i(t) = n(t) x_i(t)$. At this point we can include in the sum above the stationary and linear term, by simply defining $n_0(t) = n(t)$ and extending the sum to $i=0$. The separation of frequencies allow equation \ref{eq:modulation-model} to describe the non-stationarity as a linear combination of several contribution, each one the time domain product of the noise source with one of the modulation, and each one allowed to couple to the detector output with a different linear and stationary transfer function $\alpha_i(\tau)$.

In this framework, the non-stationary noise coupling has been reduced to a linear coupling problem. We can solve it directly in the frequency domain with an approach that follows closely the optimal a-causal Wiener filter \cite{wiener}. The residual after noise subtraction is defined as $r(t) = h(t) - \varepsilon_\mathrm{L}(t) - \varepsilon_\mathrm{NL}(t)$. For each frequency $\omega$ the optimal value of the coupling coefficients $\tilde \alpha_i(\omega)$  can be obtained by equating to zero the gradient of the Power Spectral Density (PSD) of the residual $S[r,r](\omega)$ with respect to each $\alpha_i(\omega)$
\begin{equation}\label{eq:gradient}
0 = \frac{\partial S[r,r](\omega)}{\partial \alpha_k{(\omega})} = H_k^* - \sum_{i=0}^N \alpha_i^* P_{ik}
\end{equation}
where the star denotes complex conjugation and we define the vector and matrices of cross spectral densities as follows:
\begin{eqnarray}
H_i(\omega) &=& S[n_i, h](\omega) \\
P_{ij}(\omega) &=& S[n_i, n_j](\omega)
\end{eqnarray}
Equation \ref{eq:gradient} can be solved directly for each frequency to obtain, in matrix notation:
\begin{equation}\label{eq:direct-fdomain}
\pmb\alpha(\omega) = \pmb{P}^{-1}(\omega) \pmb{H}(\omega)
\end{equation}
Equation \ref{eq:direct-fdomain} provides a direct solution to the problem of finding the optimal $\alpha_i$, in the sense of making the power spectral density of the residual as small as possible, independently for each frequency. It can be implemented in efficient ways using linear algebra numerical packages and Fast Fourier Transforms. However, this direct frequency-domain approach has several drawbacks: it is not possible to force the coupling coefficients $\alpha_i$ to be causal or stable in the Laplace sense \cite{stability} (all poles on the left half $s$-plane). Although it is still possible to perform the noise subtraction in the frequency domain, having non-physical coefficients (i.e. non-causal) can be troublesome, since past and future are mixed in the result. Moreover, each frequency is treated separately, meaning that the number of free parameters in the solution can be very large, often resulting in overfitting and oversubtraction.

To overcome the problems stated above, we can express each $\alpha_i(\omega)$ in a suitable form that uses a reduced number of parameters. A first choice could be to write each $\alpha_i$ as a rational function of order $M$ in the Laplace variable $s$. This would largely reduce the number of parameters and smooth the solutions. Overfitting would be largely reduced, but there would be no guarantee that the solutions were physically realizable in time domain, i.e. stable. To work around this problem, we use the partial fraction expansion \cite{complex-calculus}:
\begin{equation}
\alpha_k(s) = \frac{\sum_{i=0}^M b_i s^i}{\sum_{i=0}^M a_i s^i} = c + \sum_{i=1}^{2M} \frac{\rho_i}{s - s_i}
\end{equation}
The requirement that the time-domain version of each transfer function must be real implies that the poles $s_i$ and their residuals must either be real, or come in complex conjugate pairs. If we collect each complex conjugate pole pair in a second order stage ($s_i$ being the $i$-th complex pole and $\rho_i$ the corresponding complex residual), and do the same with pairs of real poles (assuming without loss of generality that there are an even number of them, where $s_{i,1}$, $s_{i,2}$ are a pair of poles with corresponding residuals $\rho_{i,1}$ and $\rho_{i,2}$), we obtain:
\begin{eqnarray}\label{eq:parametric}
\alpha_k(s) &=& c + \sum_i \frac{2\mathcal{R}[\rho_i] \, s - 2 \mathcal{R}[\rho_i^* s_i]}{s^2 - 2\mathcal{R}[s_i]\, s + |s_i|^2} \nonumber \\
&&+ \sum_i \frac{(\rho_{i,1} + \rho_{i,2}) s - (\rho_{i,1} s_{i,2} + \rho_{i,2} s_{i,1})}{s^2 - (s_{i,1} + s_{i,2}) \, s + s_{i,1} s_{i,2}}
\end{eqnarray}
where $\mathcal{R}[x]$ denotes the real part of $x$. The first sum runs over all complex pole pairs, and the second sum runs over all real pole pairs. The stability requirement can be expressed in terms of the pole position in the Laplace plane as $\mathcal{R}[s_i] < 0$ for all complex and real poles. By inspecting the form of the coefficients of the second order stages in equation \ref{eq:parametric}, we can show that the stability requirements corresponds to forcing the denominator to have strictly positive zeroth and first order coefficients. Therefore, each coupling coefficient is parametrized as
\begin{eqnarray}\label{eq:sos-parametrization}
\alpha_k(s) = c_k + \sum_{i=1}^{M/2} \frac{b^{(i)}_{k,1} s + b^{(i)}_{k,0}}{s^2 + a^{(i)}_{k,1} s + a^{(i)}_{k,0}}
\end{eqnarray}
subject to the requirements that $a^{(i)}_{k,j} > 0$ for all $i,j$ and $k$.

This parametrization solves all the problems previously mentioned concerning the frequency-domain direct solution: it drastically reduces the number of parameters, avoiding overfitting, and it ensures that the coupling coefficients $\alpha_i$ are realizable in the time domain, being causal and stable. The parametrization now mixes all frequencies, and therefore it is not possible to solve the optimization problem for each frequency independently. Instead, we need to define a scalar cost function. Considering the frequency band $\omega_L < \omega < \omega_H$ of interest for the noise subtraction, one option would be to define a cost function based on the integral of the residual PSD over that range. However, power spectral densities often have values ranging over many order of magnitudes, so this cost function could be heavily skewed toward the frequencies at which there is more signal or noise. We therefore add a frequency dependent weight function $W(\omega)$ in the PSD integral. One choice that proved to be very effective in all practical application, is to set this weight function to the inverse of the power spectral density of the detector output $W(\omega) = S[h,h]^{-1}(\omega)$. In this way the cost function takes equally into account any relative improvement on the noise, with respect to the original values. In summary, we define the cost function as
\begin{equation}\label{eq:cost-function}
C(\pmb{\theta}) = \int\limits_{\omega_L}^{\omega_H} \frac{S[r,r](\omega)}{S[h,h](\omega)}\, d\omega
\end{equation}
where $\pmb\theta = \left\{ \theta_{m} \right\}$ is a shorthand for the set of all the coupling coefficient parameters, i.e. $a, b$ and $c$ in equation \ref{eq:sos-parametrization}. Borrowing a technique commonly used in the training of deep neural networks, we can search for the minimum of the cost function by gradient descent. The gradient can be computed in closed form using the chain rule:
\begin{equation}\label{eq:gradient-params}
\frac{\partial C}{\partial \theta_m} = \int\limits_{\omega_L}^{\omega_H} \frac{1}{S[h,h](\omega)} \,   \frac{\partial S[r,r](\omega)}{\partial \alpha_k(\omega)} \frac{\partial \alpha_k(\omega)}{\partial \theta_m} \, d\omega
\end{equation}
The first partial derivative inside the integral is given by equation \ref{eq:gradient}, while the second derivative is not zero only when the index $k$ corresponds to the only $\alpha_k$ that contains the parameter $\theta_m$, and can be computed in closed form with simple algebra from the parametrization of each $\alpha_k$ given in equation \ref{eq:sos-parametrization}. 

To enforce the stability requirements, instead of carrying out a constrained optimization, we perform the following reparametrization $a^{(i)}_{k,j} \rightarrow \exp a^{(i)}_{k,j}$ so that positivity is ensured without the need of hard constraints. This reparametrization also helps compressing the coefficient dynamic range. 
With an efficient way to compute the cost function and the gradient, we can apply a gradient descent algorithm or any modification of it. By experimentation we found that the ADAM algorithm \cite{adam}, very popular for DNN training, performs very well with our optimization problem. Once the optimizer has converged to a good solution, the parameters can be easily converted back to the coefficients of Laplace domain transfer functions, or to the filter taps needed to implement a time domain IIR filter \cite{iir, bilinear}.

Different parameterizations of the coupling coefficients $\alpha_k(s)$ are possible. For example, by using a scaled sigmoid, it is possible to bound the maximum and minimum frequencies allowed for the poles. The gradient with respect to the new parameters can still be computed in a closed form.
Otherwise, we could arrange the coefficients in the denominator so that not only stability is enforced, but also both the frequency and the damping factor of all the poles are bounded, so to avoid introducing narrow resonances. Finally, we note that the $\alpha_k$ coefficients could be parametrized directly in the $z$-domain \cite{bilinear} used to describe discrete-time signals, so that we do not need to convert the Laplace-domain transfer function coefficients to time-domain, since the result of the algorithm will directly be the IIR filter taps.

One drawback of our approach is that the optimization problem is no longer linear in the parameters, and therefore there is no direct, closed-form solution. This, and the use of a gradient-based optimization algorithm, means that there is no guarantee of converging to the global optimal solution. In practice, the parametrization described above in equation \ref{eq:sos-parametrization} ensured a fast convergence in all cases studied, with performance in line with the optimal frequency-domain solution (provided there was no overfitting in the latter).

\begin{figure*}[t] 
\begin{center}
\includegraphics[width=1.02\columnwidth]{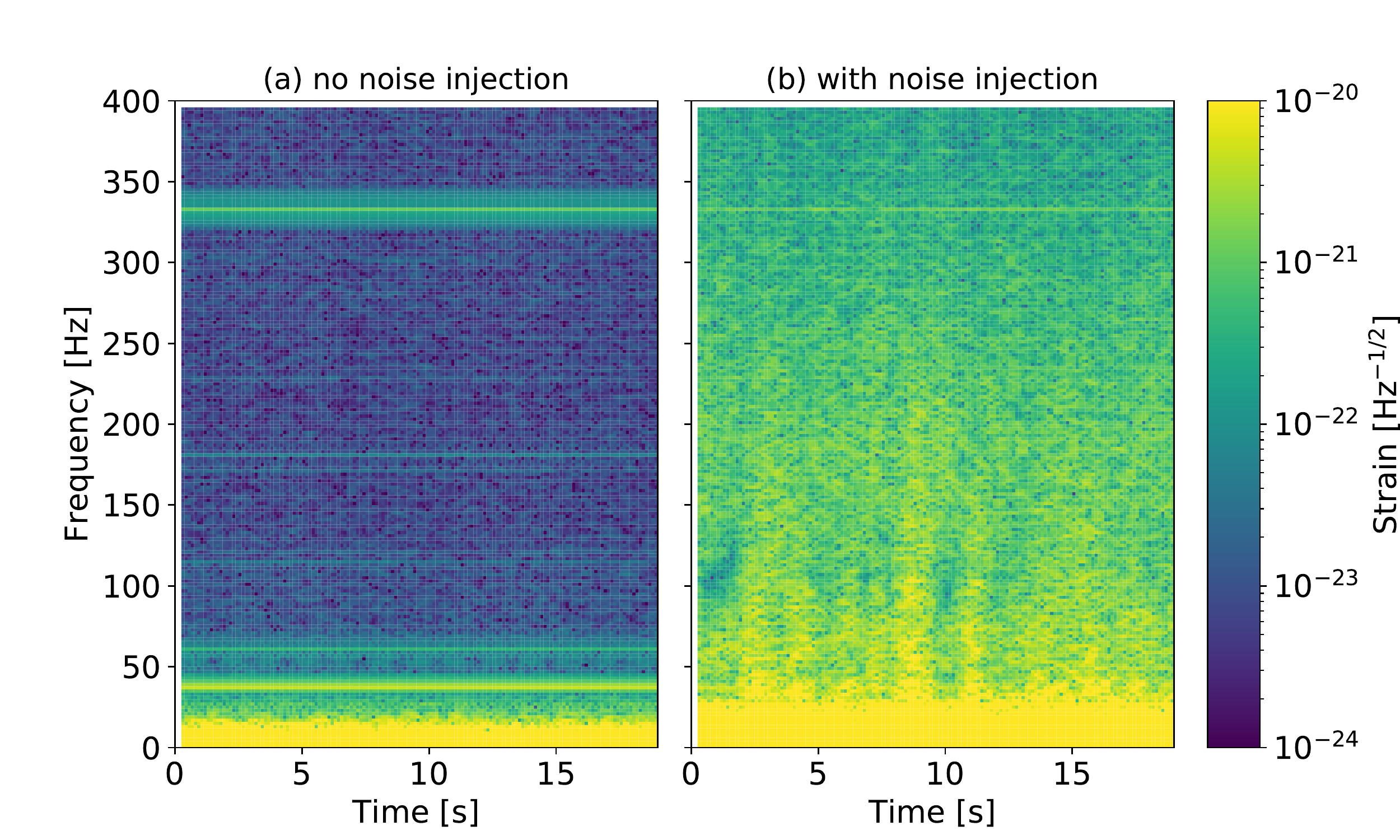} \includegraphics[width=0.9\columnwidth]{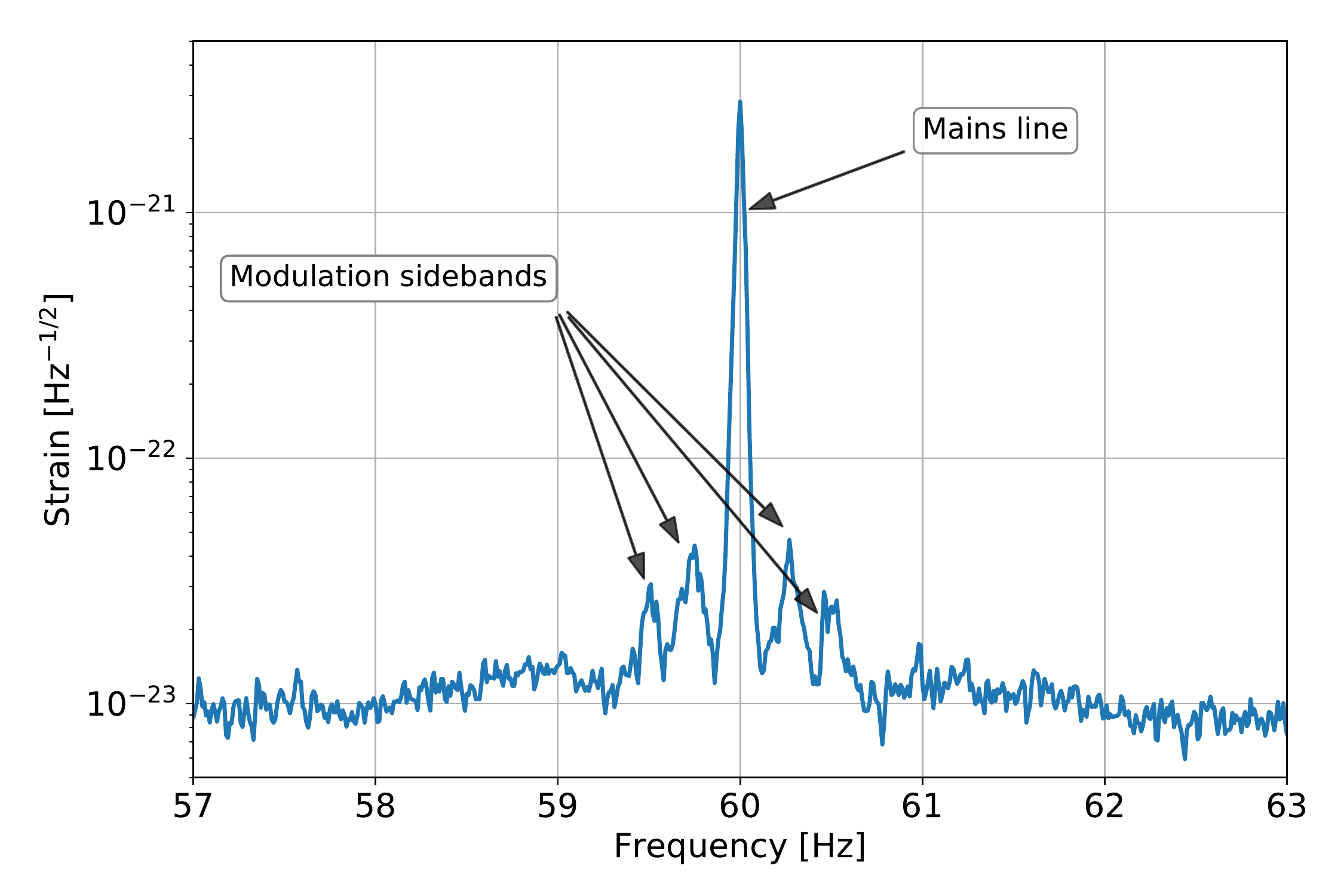} 
\end{center}
\caption{\label{fig:noise} Two examples of non-stationary noise couplings. The left panel shows a time-frequency spectrogram of the Hanford LIGO detector main output: in (a) during a quiet time of detector operation, in (b) during a period of time when random noise was purposely added to the signal recycling cavity length control. Despite the added noise being stationary over time, the effect on the detector output, between 20 and 300 Hz, changes on a time scale of the order of seconds, meaning that the noise couplings are non stationary. The right panel shows an amplitude spectral density of the LIGO Hanford detector output around the 60 Hz power line. There are symmetrical sidebands around the main frequency, evidence that the coupling of the electromagnetic noise at 60 Hz is modulated over time.}
\end{figure*}

\section{Applications to GW detector noise}\label{sec:results}

In this section, we shall consider two examples of applications of the algorithm, inspired by non-stationary noise couplings found in gravitational-wave detectors, with particular emphasis on the Advanced LIGO detectors \cite{aligo, aligo-sensitivity}. In both cases described here, the noise witnessed by an auxiliary sensor or control loop was modulated by residual angular motions of the interferometer mirrors and laser beam.  

\subsection{Signal Recycling Cavity Length noise.}

In the first example the noise source is linked to the longitudinal control system needed to keep the interferometer at its most sensitive working condition, using feedback controls that maintain all resonant cavities at the operating point \cite{length-sensing-control}. Those feedback control loops can introduce noise in the interferometer auxiliary degrees of freedom, due to their sensing or actuation limitations \cite{aligo-sensitivity}. This excess displacement noise can couple to the GW strain signal. One important example, shown in the left panel of figure \ref{fig:noise}, is related to the signal recycling cavity length (SRCL) control \cite{srcl}. Experimental evidence shows that displacement noise in this degree of freedom couples to the GW strain signal in a non-stationary way. The spectrogram in figure \ref{fig:noise} shows the detector strain while the SRCL noise was deliberately increased to enhance the effect. The noise amplitude modulation is due to residual angular motion of the interferometer mirrors around their nominal positions. There is also a linear and constant coupling coefficient, but this is partially compensated online by using a feedforward technique \cite{noise-feedforward}. 

\begin{figure}[tb] 
\begin{center}
\includegraphics[width=0.9\columnwidth]{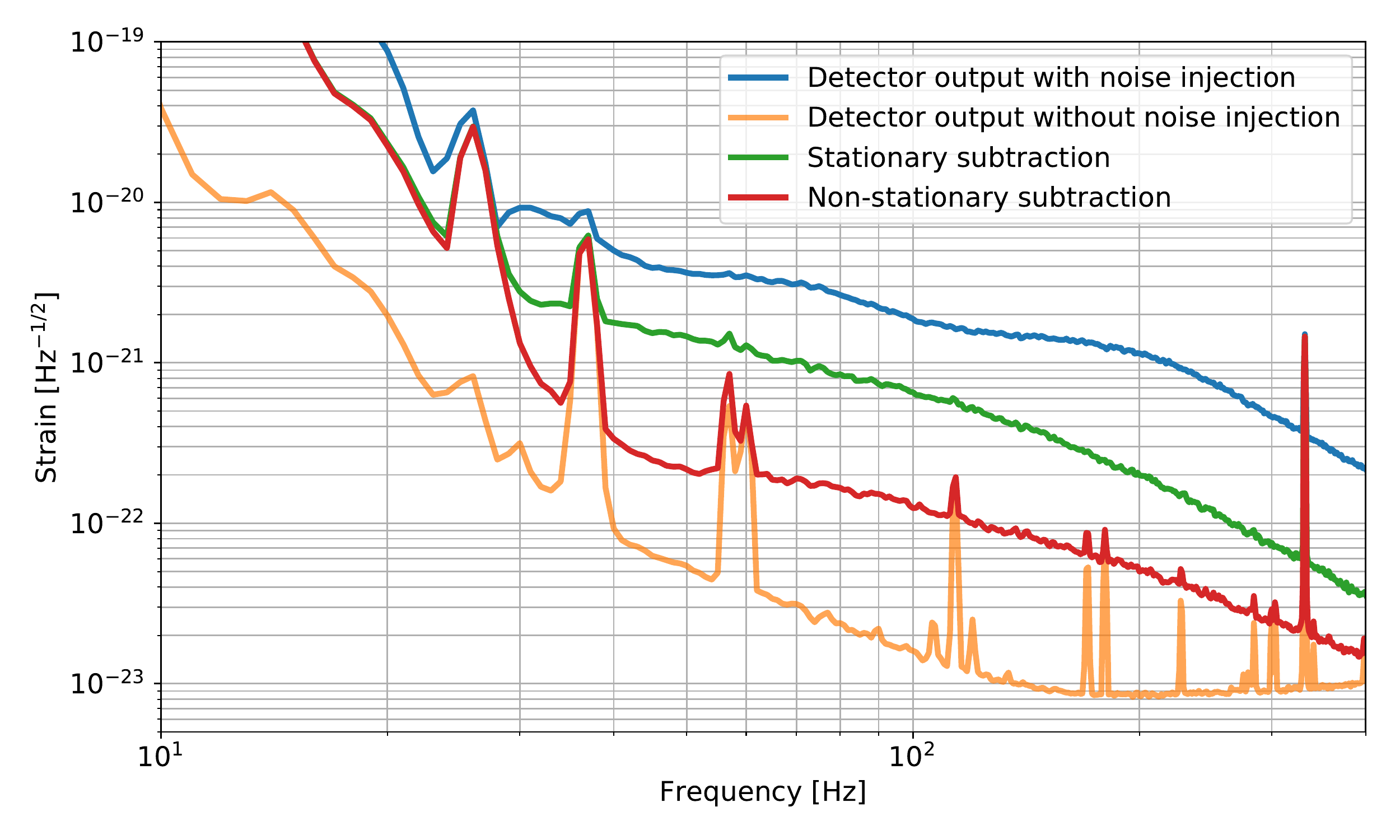} \includegraphics[width=0.9\columnwidth]{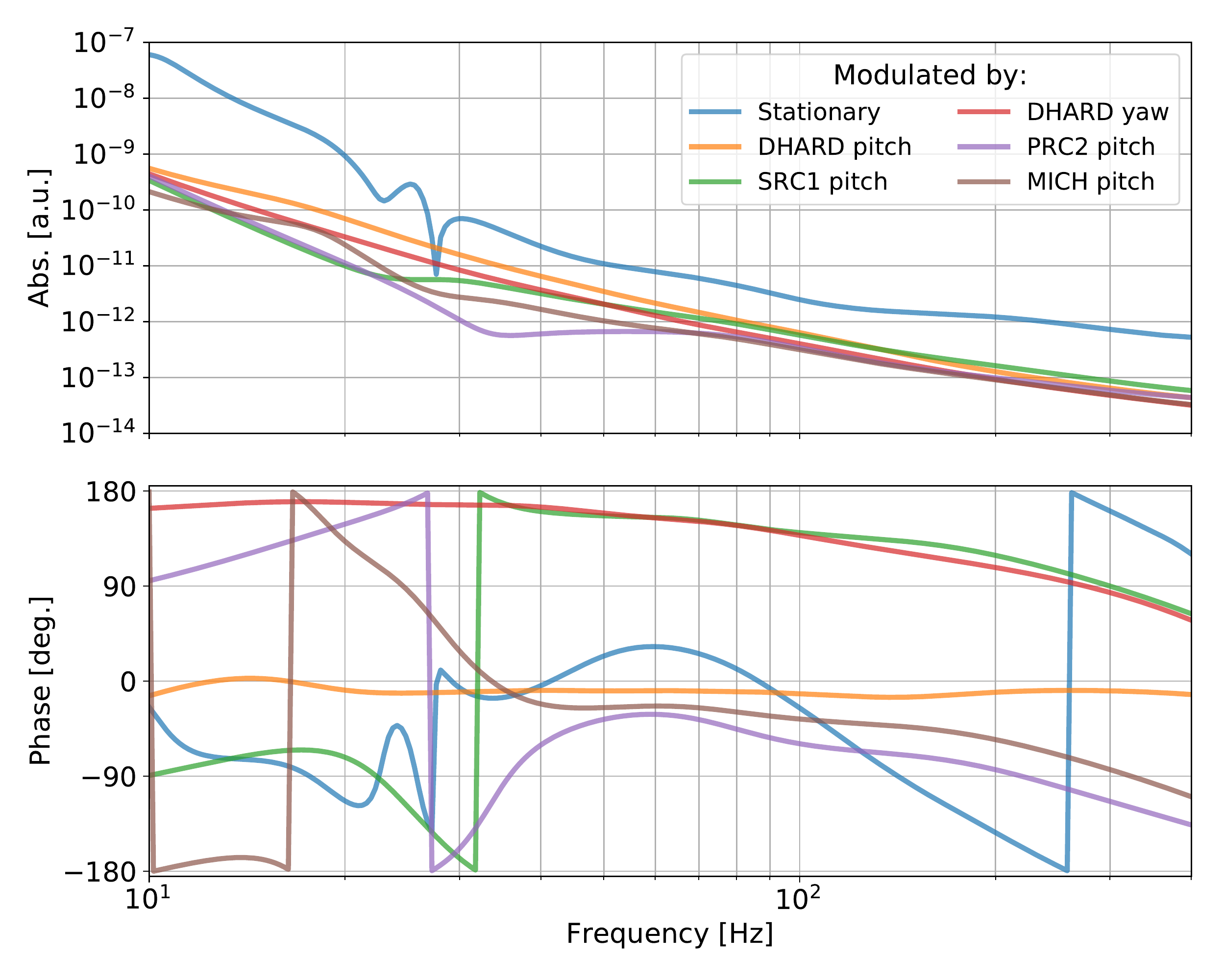} 
\end{center}
\caption{\label{fig:srcl-results} Performance of  the non-stationary noise algorithm applied to the coupling of signal recycling cavity length noise. The top panel shows the amplitude spectral density of the detector main output: the orange curve is a reference sensitivity when there was no noise injection, while the blue curve is the detector sensitivity during the noise injection and without any subtraction applied. If only the linear and stationary coupling is estimated and subtracted, the result is the green curve. By using the algorithm described here, a non-stationary subtraction gives the red curve, which largely improve upon the linear subtraction. The best non-stationary subtraction cannot remove all of the noise couplings: the reason being that the residual coupling modulation is not witnessed by the set of signals used in this work. The bottom panel shows the first few most important contributions to the modulated transfer functions $\alpha_k$ as produced by the algorithm. The largest term is the stationary transfer function, while the others are labelled with reference to the angular motion of the modulation source. For reference, DHARD is a combination of the arm cavity mirrors, moving in a differential way in the two interferometer arms \cite{sidles-sigg}; SRC1 and PRC1 denotes respectively the signal and power recycling cavity angular degrees of freedom \cite{angular}; MICH denotes the motion of the beam splitter mirror \cite{angular}.  }
\end{figure}

In this case, the noise source witness sensor $n(t)$ is the digital output of the feedback loop, sampled at a frequency of 16384 Hz. The target signal $h$ is the main detector output, which is in units of calibrated strain and sampled at 16384 Hz. Random noise was added to the SRCL control loop, to make sure that the effect dominated over the background detector output by one to two orders of magnitude. As shown in figure \ref{fig:noise}, the resulting detector output shows modulated noise. The coupling modulations $x_i$ are witnessed by the residual motion of the interferometer angular degrees of freedom, measured by the input signals to the angular feed back control systems \cite{asc}, sampled at 16 Hz. Each mirror is controlled in orientation both around the vertical axis (yaw) and around the horizontal axis perpendicular to the laser beam (pitch). Instead of controlling each mirror separately, their motions are combined in physical degrees of freedom \cite{angular, asc} that are closely related to the laser resonance conditions in the interferometer. 

The modulated signals were constructed as explained in the previous section, and each of the coupling coefficients $\alpha_k$ was parametrized as the sum of 30 second-order stages, as in equation \ref{eq:sos-parametrization}. The optimization problem consisted in the minimization of the residual signal power between 10 and 400 Hz, weighted by the inverse of the initial power spectral density, as in equation \ref{eq:cost-function}. The optimization was carried out using analytical forms for the gradient, implemented in python and accelerated using code deployed to GPU with TensorFlow \cite{tensorflow}. The optimization process took an approximate time of ten minutes on a Nvidia Titan GPU \cite{nvidia}, using 600 seconds of training data. A similar amount of data has been set aside to test the subtraction performance, and not used for parameter training.

Figure \ref{fig:srcl-results} shows the results. The algorithm output, obtained in terms of second-order stages, was converted to IIR filters subsequently implemented in the time domain. The result was then used to compute the power spectral densities shown in the figure. In the top panel, the detector sensitivity during the noise injection is compared with a reference quiet time. If only the residual linear stationary term is subtracted, for example using a Wiener filter, the noise level is reduced by less than a factor of 10 at all frequencies. The subtraction can be improved significantly at all frequencies by using the output of the non-stationary algorithm described here. The residual is not at the level of the quiet reference, meaning that the set of witness signals is not enough to capture the entirety of the modulation. The bottom panel of figure \ref{fig:srcl-results}  shows the magnitude and phase of the first few coupling coefficients $\alpha_k$, ranked by the amount of subtraction they provide. The largest contribution is the stationary term, but the first non-stationary contributions are following less than one order of magnitude below. The results show also that each modulation channel can couple to the detector output with a different transfer function, meaning that the physical coupling path is likely different. The results also show that this algorithm is capable of capturing complex and diverse frequency dependencies for each coupling path. 

The algorithm described provided a clear indication of the sources of the non-stationarity, and this information could be used to improve the detector angular stability and thus reduce the problem at the root. As a result, during normal operations of the LIGO detectors, the SRCL control is not a source of noise that limits the sensitivity, and therefore there was no need to implement the non-stationary subtraction online.

\subsection{Power line} \label{sec:60hz}

\begin{figure}[tb] 
\begin{center}
\includegraphics[width=\columnwidth]{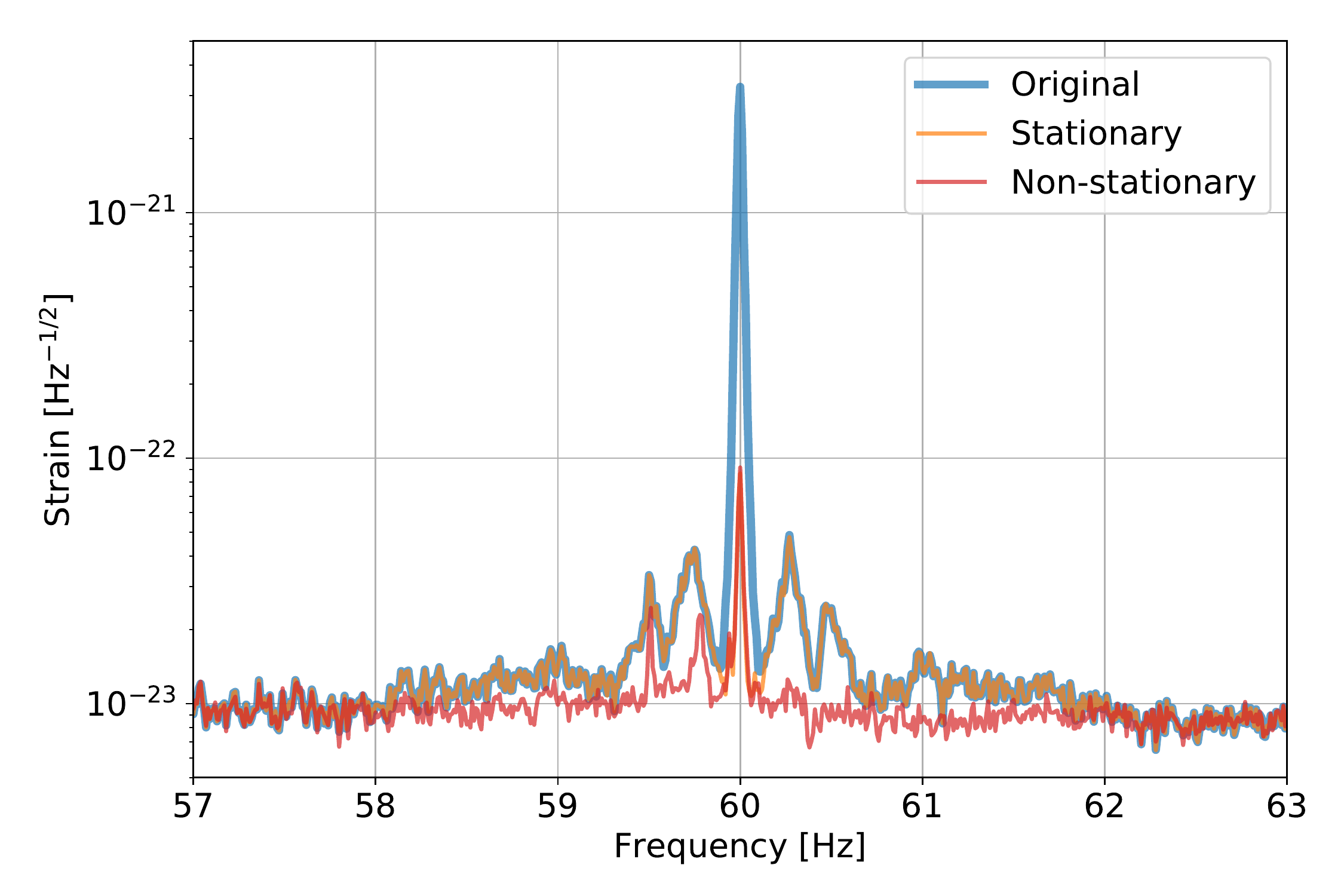}
\end{center}
\caption{\label{fig:60hz-results} Application of the non-stationary noise subtraction to the 60 Hz main power line at the Advanced LIGO Hanford detector. This figure compares the original detector output (blue) to what can be obtained by simply performing a linear and stationary subtraction of a witness channel (orange), and to the improved subtraction obtained when allowing for coupling modulation (red). The stationary subtraction matches the non-stationary one only at the 60 Hz line frequency, and has no effect at all other frequencies.}
\end{figure}

In the second example the noise source is electromagnetic in nature, and due to the 60 Hz line generated by the main power supplies. Despite many mitigation efforts, electromagnetic fields at 60 Hz couple to the detector output through many paths \cite{pem}. The linear and stationary coupling is dominant, as can be seen in the right panel of figure \ref{fig:noise}. However, the line is surrounded by symmetric sidebands that arise because the coupling is modulated by slow ($\lesssim 2$ Hz) angular motions of the interferometer beam and mirrors, similarly to the SRCL noise case.  This is another example of non-linear or non-stationary couplings. As shown in figure \ref{fig:60hz-results}, a simple linear subtraction is effective at reducing the main line by orders of magnitude (using a sensor that witnesses the power line), but leaves the sidebands untouched. This limits the detector sensitivity on a wider frequency band. This effect is significant in the Advanced LIGO Hanford detector, used in the example discussed here, and present to a lower extent in the Advanced LIGO Livingston detector.

The algorithm described in section \ref{sec:parametrized} has been applied to this problem, restricting the computation of the cost function to a narrow frequency band that includes the main line and sidebands ($50 \mbox{ Hz} < f < 70 \mbox{ Hz}$). The noise witness sensor is a direct monitor of the power supply (largely dominated by the single-frequency 60 Hz line and its harmonics). The modulation witness sensors are the same angular motion signals used in the SRCL case. Since we are subtracting noise in a narrow band around 60 Hz, we did not expect to need complicated transfer functions, so we restricted the coupling coefficients $\alpha_k$ to be modeled by only a constant plus one second-order stage:
\begin{equation}
\alpha_k(s) =  c_k +\frac{b_{k,1} s + b_{k,0}}{s^2 + a_{k,1} s + a_{k,0}}
\end{equation}
allowing us enough freedom to adjust the coupling phase and gain near 60 Hz. The result is shown in figure \ref{fig:60hz-results}: the modulated noise subtraction removes the main 60 Hz to the same level as the linear subtraction, and also reduces all the sidebands by a factor of at least 2, down to a level compatible with the surrounding background noise. 

\subsection{Effect on astrophysical range and source parameter estimation}

\begin{figure}[tb] 
\begin{center}
\includegraphics[width=0.9\columnwidth]{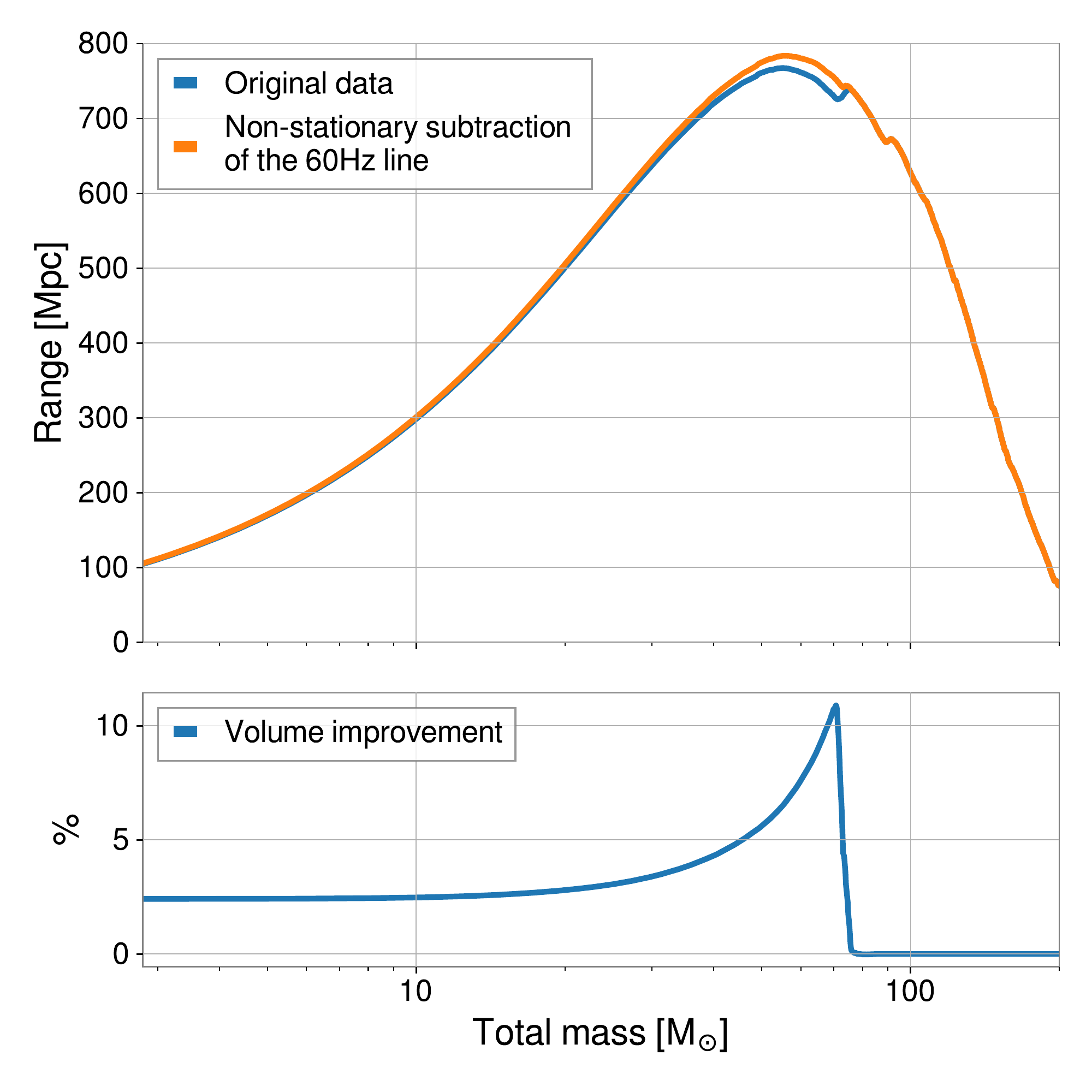}
\end{center}
\caption{\label{fig:range} The top plot shows the sky-averaged range for binary system coalescence, as a function of the total mass of the two objects. The non-stationary subtraction of the 60 Hz line and sidebands results in an improvement in the range. The range increase is small for binary neutron stars, from 104.6 to 105.5 Mpc, since the signal for those systems sweep through the 60 Hz region quickly. The improvement is more significant for large mass binary black holes, where more signal is accumulated around 60 Hz. For a total mass of 70 M$_\odot$ the range increases from 729.3 to 754.6 Mpc. The bottom plot shows the increase in observable volume as a function of the total system mass.}
\end{figure}

As discussed above, the  signal recycling cavity noise did not limit the detector sensitivity during the last period of operation. On the other hand, the non-stationary subtraction of the 60 Hz line and sidebands was effective at improving the astrophysical sensitivity of the Advanced LIGO detectors during the first six months of the O3 observation run. One way to quantify the improvement is to compute the range of the detector: the sky-averaged distance at which a compact binary coalescence can be detected with a signal-to-noise ratio of 8 \cite{range}. Figure \ref{fig:range} shows that the 60 Hz subtraction has a significant impact on the detector range for high mass binary black hole systems, increasing the detector range for systems with a total mass of 70 M$_\odot$ by 25 Mpc and the observable volume by 11\%. 

It is important to check that the non-stationary subtraction does not affect the interferometer response to GW signals and calibration. For this purpose, we applied sinusoidal forces to the interferometer test masses (focusing on a frequency range around 60 Hz), using the photon calibrator \cite{calibration, pcal}, and thus generating a differential length change in the two interferometer arms that mimics the effect of a GW. We then checked that the amplitude and phase of the calibrated detector output matched the expectation, and that the non-stationary subtraction did not affect the results, within the measurement uncertainties.

Another important check is that the non-stationary cleaning does not corrupt astrophysical signals in the data.
To corroborate this, we inject simulated binary black hole signals into linearly-cleaned strain data and then apply the additional non-stationary subtraction.
For data with and without the non-stationary subtraction, we recover the signal properties using \texttt{lalinference}, LIGO and Virgo's standard Bayesian parameter estimation infrastructure \cite{1409.7215}.
We carry out injections at two times during which contamination from the 60 Hz line was noticeable in the linearly-cleaned data from LIGO Hanford (GPS times 1244006580 and 1243309096), similar to figure \ref{fig:60hz-results}.
For each of those times, we inject signals with all combinations of three total mass values ($M = m_1 + m_2 = 200,\, 275,\, 350~M_\odot$) and two mass ratios ($q= m_2/ m_1 = 0.5,\, 1$), and always without spin in either component ($a_1=a_2=0$).
The masses are chosen so that the final cycles of the GW signal have significant frequency content in the vicinity of 60 Hz.
We additionally study a signal with $M=70\, M_\odot$ and $q=1$ at GPS time 1244006580, meant to roughly correspond to the peak of the sensitive-volume improvement in Fig.~\ref{fig:range}.
Each injection is carried out with optimal network signal-to-noise ratios (SNR) of 15 and 30\footnote{Computed using the data where the 60 Hz line was subtracted linearly.}, and into a three-detector network of two Advanced LIGO detectors and the Advanced Virgo detector. For this analysis we applied the non-stationary noise subtraction only to the Advanced LIGO Hanford detector data, since the effect on the Livingston detector was negligible.  
In all cases, the injections are produced using the numerical-relativity surrogate waveforms \texttt{NRSur7dq2} \cite{1705.07089}, and recovered with the spin-precessing waveform approximants \texttt{IMRPhenomPv2}~\cite{1308.3271}, which is standard in LIGO-Virgo analyses.
For control purposes, PSDs are estimated both through a simple Welch average \cite{10.1109/TAU.1967.1161901} and a Bayesian model using \texttt{BayesLine} \cite{1410.3852}.

\begin{figure}[tb] 
\begin{center}
\includegraphics[width=0.9\columnwidth]{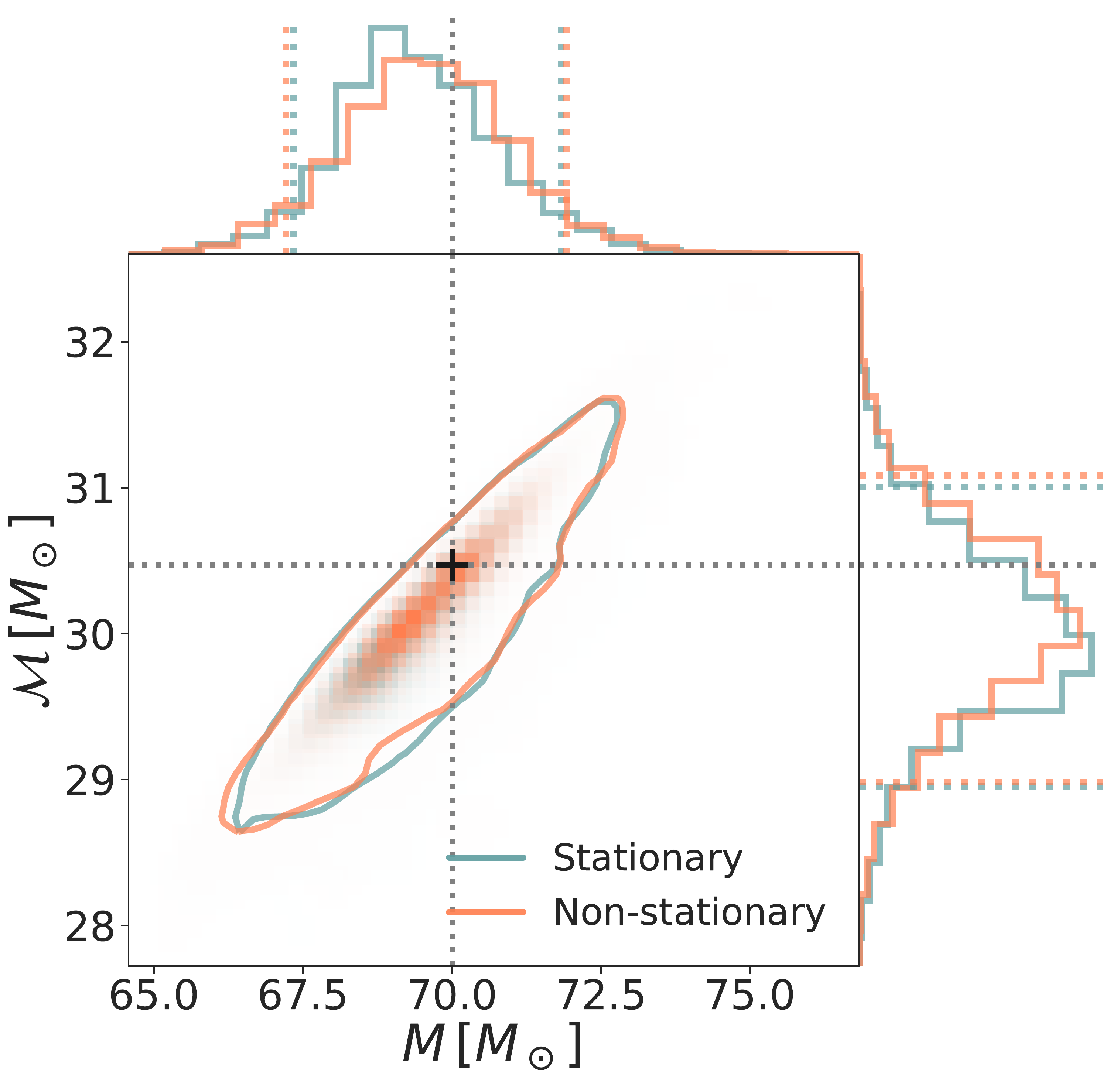}
\end{center}
\caption{\label{fig:pe} Joint posterior probability density on the total mass $M=m_1 + m_2$ ($x$-axis) and chirp mass $\mathcal{M} = (m_1 m_2)^{3/5} M^{-1/5}$ ($y$-axis), for the $M=70\,M_\odot$ and $q=1$ injection with SNR 30, at GPS time 1244006580, recovered using a Welch-average estimate of the noise PSD. Colors correspond to the non-stationary cleaning of the data (orange) and to the linear cleaning of the data (blue). The main panel shows the 2D probability density, with solid contours containing 90\% of the probability mass. The secondary panels above and to the right show the corresponding 1D marginalized distributions for $M$ and $\mathcal{M}$ respectively, with colored dashed lines representing symmetric 90\%-credible intervals. The true values are marked by a crosshair and gray dotted lines.}
\end{figure}

Our results indicate that the non-stationary subtraction does not adversely impact parameter estimation and, therefore, does not corrupt astrophysical signals in the data. The lack of discernible improvement after the non-stationary cleaning is expected given that, in this case, only the Hanford detector data was affected, and the Livingston detector was the most sensitive in the network at that time.
As an example of this, Figure \ref{fig:pe} shows the recovered posterior distributions of the system's total mass $M$ ($x$-axis) and chirp mass $\mathcal{M} = (m_1 m_2)^{3/5} M^{-1/5}$ ($y$-axis), for the $M=70\, M_\odot$ and $q=1$ injection at GPS time 1244006580.
The result for the two cleaning techniques (linear and non-stationary) are not significantly different. However, the non-stationary step improves the recovered matched-filter SNR by a factor consistent with the range improvement displayed in Fig.~\ref{fig:range}.
This seems to be the case for all recovered parameters and for all of the injections in our set.

\section{Extensions and other applications}\label{sec:extra}

The algorithm presented here was inspired by the non-stationary noise couplings found in gravitational wave detectors, where a noise source with power in the tens to hundreds Hz region can limit the detector sensitivity, and be modulated by slower (below a few Hz) residual motions. However, the parametrized approach to the noise subtraction can be extended to any other application when there is a noise coupling which is modulated. It can also be extended to the case of quadratic or higher order couplings, even when there is no clear separation of the signal frequency support. This is possible by choosing a set of noise witness sensors $w_i$, constructing the set of all quadratic (or higher order) combinations $n_{ij} = w_i w_j$ and using them in equation \ref{eq:modulation-model}.

The parametrization described above for the coupling coefficients turns out to be quite versatile and robust. Even when considering only linear and stationary noise couplings, the algorithm described here is able to match the performance of the  frequency-domain Wiener filter. It is therefore a viable approach to a stable and causal Wiener filter that can be implemented in time domain using IIR filters. The advantage over the classical finite impulse response (FIR) Wiener filter \cite{wiener} is the significant reduction in the number of parameters, the lower computational cost of the time domain implementation, and the absence of overfitting problems. The main drawback is again that there is no closed form solution, and the parameters must be found by a gradient-descent-based optimization, with no guarantee of optimality.

In the treatment of non-stationary noise described above, we assumed that the change in the couplings could be completely captured by a set of modulation witness signals. This might not always be the case. The set of witness signals might be incomplete, resulting in some residual modulation at the same time scale as those that are modeled and removed. This was the case in the residual noise coupling for the SRCL noise, as shown in figure \ref{fig:srcl-results}. Another possibility is that the set of modulation signals is sufficient to describe the non-stationarity for a short period of time, still longer than the modulations witnessed by the signals, but the coupling coefficients $\alpha_k$  vary on a time scale which is slower than the typical content of the witness channels. In this case we would need to slowly adjust the parameters of the noise subtraction. In equation \ref{eq:gradient} we expressed the gradient of the cost function with respect to the parameters in a form that is numerically efficient to find the optimal parameters, since the cross spectral densities need to be computed only once at the beginning of the training. However, if the noise couplings change over time, it is more convenient to rewrite the gradient in the following form:
\begin{equation}
\frac{\partial S_{r,r}(\omega)}{\partial \alpha_k(\omega)} = - S[r, n_k]
\end{equation}
that can be obtained with straightforward manipulations of equation \ref{eq:gradient}. The gradient can be computed by accumulating the (varying) cross spectral densities of the current subtraction residual with all the modulation signals. This gradient can then be applied to the minimization of a running cost function as in equation \ref{eq:gradient-params}, with an approach similar, for example, to the least mean squared (LMS) algorithm \cite{lsm}.
 
\section{Conclusions}\label{sec:conclusions}

We presented a novel algorithm to characterize and subtract non-stationary noises from the output signals of physical detectors, which can be applied to all cases when one or more fast noise sources are coupling to the main detector output via modulated transfer functions. Provided there is access to suitable witness sensors that track both the noise and the modulations, we show how a parametrized, stable and time-domain noise cancelation can be implemented. This extends the well-known noise cancellation techniques based on feedforward and Wiener filters, and allow for a real-time implementation of non-stationary noise subtraction.

We show how this technique can be applied successfully to the output of GW detectors, with examples from the Advanced LIGO observatory. The implementation of non-stationary noise subtraction allows us to improve the detector sensitivity,  because the average power spectral density of the noise is reduced below what is attainable with simple linear noise cancellations, and also because the residual is more stationary and therefore better suited to searches for GW triggers. We also show that the non-stationary noise subtraction can improve the sky-averaged detectable range, and does not introduce any bias in the astrophysical parameters estimated for simulated GW events that contain a significant amount of signal power around 60 Hz.

Finally, we note that the technique described here is of general interest, and can be applied in all cases where non-stationary noise couplings are present in any detector. It is also possible to limit the scope of the algorithm to the linear and stationary case, providing a new approach to the optimization and implementation of efficient Wiener filters. In both the stationary or non-stationary cases, it is also possible to convert this algorithm into an adaptive system, where the noise cancellation parameters vary slowly to cope with changes in the noise couplings.

\appendix

\section{Deep learning-based subtraction}\label{sec:deep-learning}

\begin{figure}[tb] 
\begin{center}
\includegraphics[width=\columnwidth]{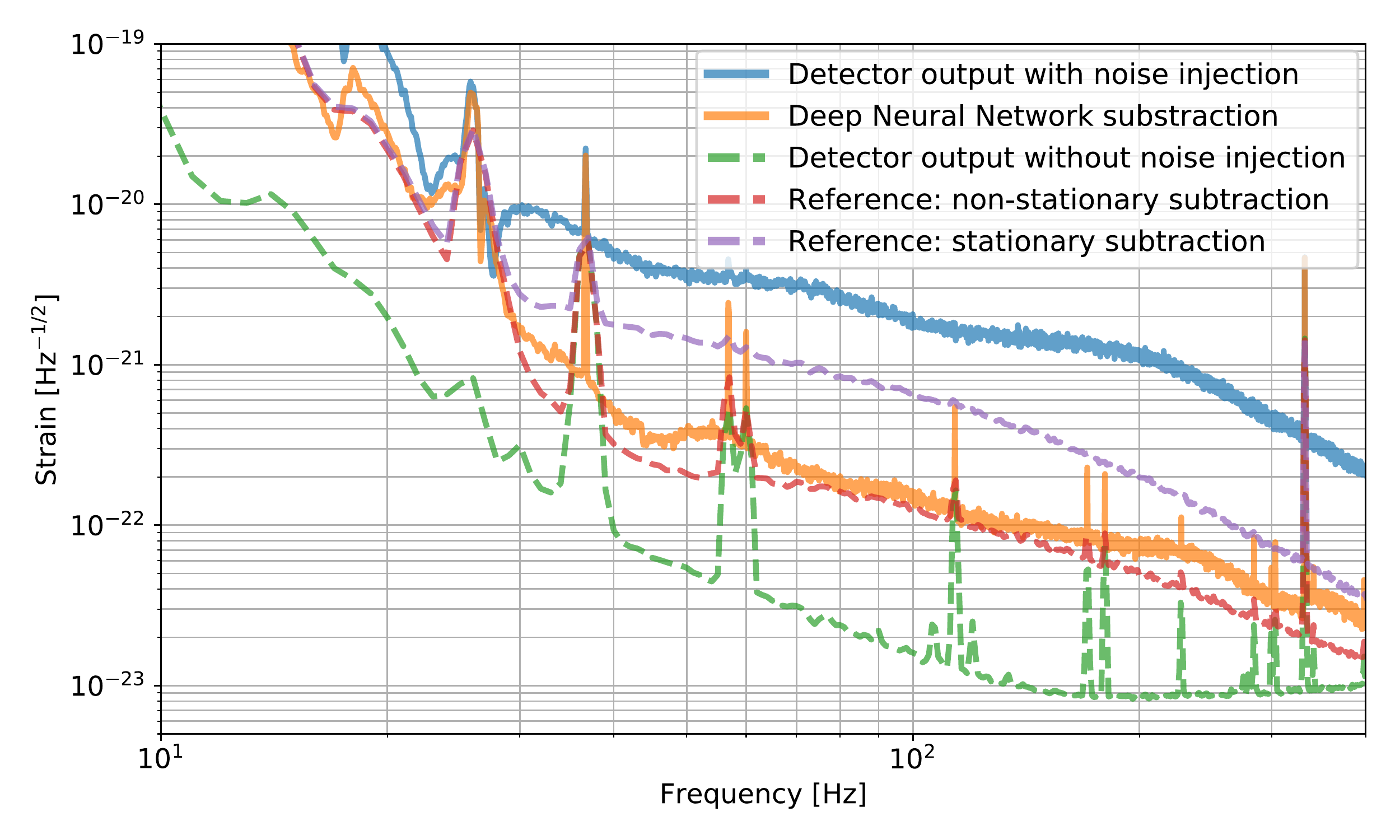}
\end{center}
\caption{\label{fig:dnn-results} Noise subtraction obtained with a Deep Neural Network, to be compared with the non-stationary noise subtraction obtained with the algorithm described in section \ref{sec:parametrized} and shown also in figure \ref{fig:srcl-results}.  }
\end{figure}

Neural networks are not a new idea \cite{perceptron}, but have gained momentum in the recent years with the application of Deep Neural Networks (DNN) \cite{dnn} to many Machine Learning problems. Ideally, a neural network is capable of approximating any non-linear function of its inputs, provided it includes a large enough number of basic units or neurons \cite{dnn-approximators}. Therefore a DNN seems to be a suitable starting point for a parametrization of the non-linear coupling function introduced in equation \ref{eq:full-non-linear}. Since the noise subtraction problem deals with processing and reconstructing time series, it is important that the DNN includes some memory of the past inputs. For this reason our attention focused on recurrent neural networks (RNN) \cite{rnn}. Despite the intrinsic non-linearity of each layer, a DNN is not particularly suitable to learn efficiently multiplications of its inputs. Since this is an important operation for most of the noise subtraction schemes we are considering, we added an ad-hoc quadratic layer: the $n$ inputs to the layer are multiplied pair-wise to obtain $n^2$ new signals; together with the input, those $n^2+n$ signals are then passed through a fully connected layer to reduce the dimensionality to $m < n^2+n$. This additional layer, preceded or followed by additional recurrent layers, largely improve the learning speed of a DNN.

We applied a DNN to the problem of subtracting the signal recycling noise described in section \ref{sec:results}. The architecture consists of three layers of Gated Recurrent Units (GRU) \cite{gru} with 64, 32 and 16 neurons each. The input to the recurrent layers consists of both the fast noise witness (signal recycling longitudinal servo output) and the up-sampled modulation witnesses (angular signals). The output of the three recurrent layers is then fed to the quadratic layer described above, and then to three fully-connected layers with 16, 16 and 8 units with ReLU activation \cite{relu}. The final signal is obtained by linearly combining the outputs of the last layer. The network has about 9000 parameters that are trained using a standard ADAM algorithm on the mean square error of the output with respect to the desired signal (the detector strain). The cost function was actually computed in the frequency domain, by integrating the residual between 10 and 400 Hz (similar to what explained in section \ref{sec:parametrized} and equation \ref{eq:cost-function}). The network was implemented in PyTorch and trained using 600 seconds of data on the same Nvidia GPU used for the main results described in this paper. The training required about 10 hours. The best subtraction obtained with this network is shown in figure \ref{fig:dnn-results}, compared with the output from the non-stationary noise subtraction algorithm described in this paper. The performance of the network is clearly better than a simple linear and stationary subtraction, but falls short of what is achievable with the non-stationary subtraction algorithm described in this work. Additionally, it is extremely difficult, if not impossible, to extract useful information from the trained network, such as what signals are the worst offenders for the non-stationarity of the couplings. It is in other words not possible to produce the equivalent of the bottom panel of figure \ref{fig:srcl-results}, therefore missing crucial information that could be used to solve the modulation problem at the root.

\section*{Acknowledgments}
The authors would like to thank C.J. Haster for help generating the simulated sources described in Sec IV.3. LIGO was constructed by the California Institute of Technology and Massachusetts Institute of Technology with funding from the National Science Foundation and operates under cooperative agreement PHY-1764464. M.I.\ is supported by NASA through the NASA Hubble Fellowship grant No.\ HST-HF2-51410.001-A awarded by the Space Telescope Science Institute, which is operated by the Association of Universities for Research in Astronomy, Inc., for NASA, under contract NAS5-26555. MJS is supported by the National Science Foundation Grant PHY 1806165. The authors are grateful for computational resources provided by the LIGO Laboratory and supported by the National Science Foundation Grants PHY-0757058 and PHY-0823459. This paper has LIGO document number P1900335.



\begin{thebibliography}{10}

\bibitem{wiener} Norbert Wiener, \emph{Extrapolation, Interpolation, and Smoothing of Stationary Time Series}. New York 1949, John Wiley \& Sons. ISBN 0-262-73005-7

\bibitem{adaptive} S. S. Haykin, \emph{Adaptive Filter Theory}, Prentice Hall (2002)

\bibitem{aligo}
J~Aasi and others [{LIGO Scientific Collaboration}].
\newblock Advanced {LIGO}.
\newblock {\em Class. Quantum Grav.}, 32:074001, 2015.

\bibitem{kagra}
T. Akutsu and others [{KAGRA collaboration}].
\newblock KAGRA: 2.5 generation interferometric gravitational wave detector
\newblock Nat Astron 3, 35?40 (2019)

\bibitem{avirgo}
F~Acernese and others [{Virgo Collaboration}].
\newblock {Advanced Virgo}: a 2nd generation interferometric gravitational wave
  detector.
\newblock {\em Class. Quantum Grav.}, 32:024001, 2015.

\bibitem{geo600}
B.~Willke, B. et al. 
\newblock The {GEO 600} gravitational wave detector. 
\newblock {Class. Quantum Grav.} 19, 1377?1387 (2002)


\bibitem{gw-catalog} B.P.Abbott and others [{LIGO Scientific Collaboration} and {Virgo Collaboration}]. 
\newblock {GWTC-1: A Gravitational-Wave Transient Catalog of Compact Binary Mergers Observed by LIGO and Virgo during the First and Second Observing Runs}.
\newblock arXiv:1811.12907 [{astro-ph.HE}]

\bibitem{quantum-noise} Danilishin, S.L. and Khalili, F.Y. 
\newblock  Quantum Measurement Theory in Gravitational-Wave Detectors.
\newblock Living Rev. Relativ. (2012) 15: 5. https://doi.org/10.12942/lrr-2012-5

\bibitem{thermal-noise} Harry, G.M., Gretarsson, A.M., Saulson, P.R., and others.
\newblock Thermal noise in interferometric gravitational wave detectors due to dielectric optical coatings. 
\newblock Classical and Quantum Gravity, 19(5), (2002) p.897

\bibitem{newtonian-noise} Hughes, S.A. and Thorne, K.S.
\newblock Seismic gravity-gradient noise in interferometric gravitational-wave detectors. 
\newblock Physical Review D, 58(12), (1998) 122002.

\bibitem{aligo-sensitivity} Martynov, D. V. and others.
\newblock Sensitivity of the Advanced LIGO detectors at the beginning of gravitational wave astronomy
\newblock {\em Phys. Rev.} D 93:112004 (2016)

\bibitem{detchar-glitches} B. P. Abbot et al. [LIGO Scientific Collaboration, Virgo Collaboration], \emph{Effects of data quality vetoes on a
search for compact binary coalescences in Advanced LIGO's first observing run}, Class Quant Grav. {\bf 35}, 065010 (2018)

\bibitem{gstlal} C. Messick et al., \emph{Analysis framework for the prompt discovery of compact binary mergers in gravitational-wave data}, Phys. Rev. D {\bf 95}, 042001 (2017)

\bibitem{pycbc} S. A. Usman et al., \emph{The PyCBC search for gravitational waves from compact binary coalescence}, Class. Quantum Grav. {\bf 33}, 215004 (2016)

\bibitem{match-filtering} C. W. Helstrom, \emph{Statistical Theory of Signal Detection}, 2nd edition (Pergamon Press, London, 1968)

\bibitem{driggers2018} J. C. Driggers et al, \emph{Improving astrophysical parameter estimation via offline noise subtraction for Advanced LIGO}, arXiv:1806.00532 [astro-ph.IM]

\bibitem{davis2018} D. Davis et al, \emph{Improving the Sensitivity of Advanced LIGO Using Noise Subtraction}, arXiv:1809.05348 [astro-ph.IM]

\bibitem{wiener-iir} T. E. Tuncer, \emph{Causal and stable FIR-IIR Wiener filters}, IEEE Workshop on Statistical Signal Processing, p. 118 (2003)

\bibitem{dnn} Ian Goodfellow, Yoshua Bengio and Aaron Courville, \emph{Deep Learning}, MIT Press 2016, \url{http://www.deeplearningbook.org}

\bibitem{dnn-approximators} Kurt Hornik  \emph{Approximation Capabilities of Multilayer Feedforward Networks}, Neural Networks, {\bf 4}, 251 (1991)

\bibitem{black-box} Ravid Shwartz-Ziv and Naftali Tishby, \emph{Opening the Black Box of Deep Neural Networks via Information}, arXiv:1703.00810 [cs.LG]

\bibitem{volterra} Vito Volterra, \emph{Sopra le funzioni che dipendono da altre funzioni. III}, Italy 1887: Rendiconti della Reale Accademia dei Lincei. pp. 97-105

\bibitem{stability} J. Doyle, B. Francis and A. Tannenbaum, \emph{Feedback Control Theory}, Macmillan Publishing Co., 1990

\bibitem{complex-calculus} Rudin, W., \emph{Real and Complex Analysis}, 3rd ed. McGraw-Hill, (1986)

\bibitem{adam} Diederik P. Kingma and Jimmy Ba, \emph{Adam: A Method for Stochastic Optimization}, 	arXiv:1412.6980 [cs.LG]

\bibitem{iir} Mitra, S. K., \emph{Digital Signal Processing: A Computer-Based Approach}. New York, NY: McGraw-Hill. (1998)

\bibitem{bilinear} A. Oppenheim, \emph{Discrete Time Signal Processing Third Edition}. Upper Saddle River, NJ: Pearson Higher Education, Inc. p. 504 (2010)

\bibitem{length-sensing-control} P. Fritschel et al., \emph{Readout and control of a power-recycled interferometric gravitational-wave antenna}, Appl. Opt. 40, 4988?4998 (2001)

\bibitem{srcl} J. Mizuno et al., \emph{Resonant sideband extraction: a new configuration for interferometric gravitational wave detectors}, Phys. Lett. A 175, 273?276 (1993)

\bibitem{noise-feedforward} F. Acernese, et al., Virgo Collaboration \emph{Performances of the Virgo interferometer longitudinal control system}, Astropart. Phys., 33 (2010), pp. 75-80

\bibitem{asc} K. L. Dooley et al., \emph{Angular control of optical cavities in a radiation-pressure-dominated regime: the Enhanced LIGO case}, J. of the Opt. Society of Am. A {\bf 30}, 2618 (2013)

\bibitem{sidles-sigg} J. A. Sidles and D. Sigg, \emph{Optical torques in suspended Fabry Perot interferometers}, Phys. Lett. A {\bf 354}, 167 (2006)

\bibitem{angular} L. Barsotti et al., \emph{Alignment sensing and control in advanced LIGO}, Class. Quantum Grav. {\bf 27}, 084026 (2010)

\bibitem{tensorflow} https://www.tensorflow.org

\bibitem{nvidia} https://www.nvidia.com/en-us/

\bibitem{pem} A. Effler et al. \emph{Environmental influences on the LIGO gravitational wave detectors during the 6th science run}, Class. Quantum Grav. 32 (2015)  035017

\bibitem{range} B. Allen, et al. \emph{FINDCHIRP: An algorithm for detection of gravitational waves from inspiraling compact binaries}, Phys. Rev. D 85, 122006 (2012)

\bibitem{calibration} B. P. Abbot et al., \emph{Calibration of the Advanced LIGO detectors for the discovery of the binary black-hole merger GW150914}, Phys. Rev. D 95, 062003 (2017)

\bibitem{pcal} S. Karki, et al. \emph{The Advanced LIGO photon calibrators}, Review of Scientific Instruments 87.11 (2016): 114503

\bibitem{1409.7215} J. Veitch et al., \emph{Parameter estimation for compact binaries with ground-based gravitational-wave observations using the LALInference software library}, Phys. Rev. D 91, 042003 (2015)

\bibitem{1705.07089} J. Blackman et al., \emph{A Numerical Relativity Waveform Surrogate Model for Generically Precessing Binary Black Hole Mergers}, Phys. Rev. D 96, 024058 (2017)

\bibitem{1308.3271} M. Hannam et al., \emph{A simple model of complete precessing black-hole-binary gravitational waveforms}, Phys. Rev. Lett. 113, 151101 (2014)

\bibitem{10.1109/TAU.1967.1161901} P. D. Welch, \emph{The use of fast Fourier transform for the estimation of power spectra: A method based on time averaging over short, modified periodograms},  IEEE Transactions on Audio and Electroacoustics 15, 70 (1967)

\bibitem{1410.3852} T. B. Littenberg and N. J. Cornish, \emph{BayesLine: Bayesian Inference for Spectral Estimation of Gravitational Wave Detector Noise}, Phys. Rev. D 91, 084034 (2015)

\bibitem{lsm} S. Haykin and B. Widrow editors, \emph{Least-Mean-Square Adaptive Filters}, Wiley-Interscience, Hoboken, NJ, 2003

\bibitem{perceptron} F. Rosenblatt, \emph{The Perceptron--a perceiving and recognizing automaton}. Report 85-460-1, Cornell Aeronautical Laboratory (1957)

\bibitem{rnn} S. Hochreiter and J. Schmidhuber, \emph{Long Short-Term Memory}, Neural Computation. 9 (8): 1735 (1997)

\bibitem{gru} K. Cho, B. van Merrienboer, D. Bahdanau, and Y. Bengio. "On the properties of neural machine translation: Encoder-decoder approaches". arXiv preprint arXiv:1409.1259, 2014

\bibitem{relu} R. H. Hahnloser et al., \emph{Digital selection and analogue amplification coexist in a cortex-inspired silicon circuit}, Nature {\bf 405}, 947 (2000)

\end{thebibliography}
\end{document}